\documentclass[arXiv]{actapoly}

\usepackage{siunitx}

\begin{document}

\title[Simple models of three coupled $\mathcal{PT}$-symmetric wave guides]
{Simple models of three coupled $\mathcal{PT}$-symmetric wave guides allowing
  for third-order exceptional points}

\author[J. Schnabel]{Jan Schnabel}{itp1}
\correspondingauthor[H. Cartarius]{Holger Cartarius}{itp1}
{Holger.Cartarius@itp1.uni-stuttgart.de}
\author[J. Main]{J\"org Main}{itp1}
\author[G. Wunner]{G\"unter Wunner}{itp1}
\author[W. D. Heiss] {Walter Dieter Heiss}{st,nithep}

\institution{itp1}{1. Institut f\"ur Theoretische Physik, Universit\"at
  Stuttgart, 70550 Stuttgart, Germany}
\institution{st}{Department of Physics, University of Stellenbosch,
  7602 Matieland, South Africa}
\institution{nithep}{National Institute for Theoretical Physics (NITheP),
  Western Cape, South Africa}

\begin{abstract}
  We study theoretical models of three coupled wave guides with a
  $\mathcal{PT}$-symmetric distribution of gain and loss. A realistic matrix
  model is developed in terms of a three-mode expansion. By comparing with a
  previously postulated matrix model it is shown how parameter ranges with
  good prospects of finding a third-order exceptional point (EP3) in an
  experimentally feasible arrangement of semiconductors can be determined. In
  addition it is demonstrated that continuous distributions of exceptional
  points, which render the discovery of the EP3 difficult, are not only a
  feature of extended wave guides but appear also in an idealised model of
  infinitely thin guides shaped by delta functions.
\end{abstract}


\maketitle
\section{Introduction}
\label{sec:introduction}

It is a well-known fact that the spectra of non-Hermitian quantum systems can
exhibit exceptional points of second order (EP2), i.e.\ branch point
singularities at which two eigenstates coalesce \cite{Kato,Heiss12,Moiseyev}.
They have been extensively studied theoretically \cite{Heiss90,Heiss2000,%
  Hernandez2006,Lefebvre2009,Cartarius09,Lee09,Longhi10,Guo09,Cartarius2011b,
  Gutoehrlein13,Wiersig2014a,HeissWunner15,Schwarz2015a,Menke2016a} and
their physical relevance has been demonstrated in impressive experiments
\cite{Philipp2000,Dembowski01,Dembowski2003,Dietz2007,Stehmann2004,%
  Lawrence2014a,Gao2015a,Doppler2016a,Xu2016a}.

In much rarer cases exceptional points of higher order (EP$N$) are discussed
\cite{Graefe08,HeissChiral,Graefe12,HeissWunner16}. In a matrix representation
they can be identified by the fact that the matrix is not diagonalisable.
With a similarity transformation one can reduce the matrix to a Jordan normal
form, where the EP$N$ appears as an $N$-dimensional Jordan block
\cite{Guenther07}. In a third-order exceptional point (EP3) three states
coalesce in a cubic-root branch point singularity, which already turned out to
exhibit new effects beyond those of EP2s such as  an unusual chiral behaviour
\cite{HeissChiral}. The exchange behaviour of the eigenstates for circles
around an EP3 shows a complicated structure. It does not in all cases uncover
the typical cubic-root behaviour \cite{Graefe12,Heiss2016}.

Of special interest in the investigation of exceptional points are
$\mathcal{PT}$-symmetric systems, i.e.\ systems whose Hamiltonians are
invariant under the combined action of the parity operator $\mathcal{P}$ and
the time reversal operator $\mathcal{T}$ \cite{Bender98}. In these systems the
exceptional point marks a quantum phase transition, in which real eigenvalues
merge under variation of a parameter and become complex if the parameter is
varied further in the same direction. The eigenstates of the complex
eigenvalues are not $\mathcal{PT}$ symmetric, this is only the case for the
eigenstates with real eigenvalues. One speaks of broken $\mathcal{PT}$
symmetry, and the EP marks the position of the $\mathcal{PT}$ symmetry
breaking. Since the occurrence of exceptional points is a generic feature of
the $\mathcal{PT}$ phase transition a large number of works exists
for $\mathcal{PT}$-symmetric quantum mechanics \cite{Bender1999,Znojil1999,%
  Jakubsky05,Graefe08,Jones2008a,Mehri-Dehnavi2010a,Jones2010,Graefe2012a,%
  Heiss13a,Dast14a,Abt2015a,GutoehrleinSchnabel15,Dast2016a,Kreibich14a,%
  Znojil2017a,Schwarz2017a,Klett2017a}, quantum field theories
\cite{Bender2012,Mannheim2013}, electromagnetic waves \cite{El-Ganainy2007,%
  Makris2008,Mostafazadeh2009a,Bender13,Bittner2014a,Mostafazadeh2013a,%
  Peng2014a}, and electronic devices \cite{Schindler2011}.

In these papers exceptional points of second order have been investigated in
great detail. $\mathcal{PT}$-symmetric optical wave guides, in particular, with
an appropriate coupling between them, are ideally suited to generate
higher-order exceptional points \cite{Graefe12}. Klaiman et al.\
\cite{Klaiman08} showed that a detailed theoretical modelling of a setup of
two wave guides predicts the occurrence of an EP2, and directly proved it via
its signatures, among them an increasing beat length in the power distribution.
Exactly this strategy has later been used experimentally \cite{Rueter10}.
Encouraged by these findings the model was extended by Heiss and Wunner, who
added a third wave guide between those with gain and loss to allow for a
third-order exceptional point. They studied a simplified model consisting of
infinitely thin wave guides modelled by delta functions \cite{HeissWunner16}.
In a follow-up paper a detailed investigation of a spatially extended setup
with experimentally accessible parameters was used \cite{Schnabel2017a}. It
was shown that the system is well capable of manifesting a third-order
exceptional point in an experimentally feasible procedure.

The purpose of this paper is to show that the essential properties of the
system studied in \cite{Schnabel2017a} can already be found in much simpler
descriptions, allowing for deeper insight. The whole three wave-guide setup
can be mapped to a matrix model. In such a matrix model an EP3 can be found
in a simple manner. However, the largest benefit is in the predictability of
matrix structures allowing for an easy access to an EP3. Since the influence
of the physical parameters on the matrix elements is known from the mapping,
the matrix can guide the search for appropriate physical parameter ranges.

In \cite{Schnabel2017a} it was demonstrated that the EP3 of interest is
surrounded by continuous distributions of EP2s or EP3s in the space of the
physical parameters. This effect in combination with the fact that an EP3 can
show a square-root behaviour for parameter space circles \cite{Graefe12,%
  Heiss2016} renders its identification difficult. In this paper we show that
this difficulty can be studied in the much simpler delta-functions model
introduced in \cite{HeissWunner16}.

The remainder of the paper is organised as follows. In Section \ref{sec:system}
we provide a brief introduction into the system. The matrix model is
developed in Section \ref{sec:matrixmodel}, where we introduce the mapping of
the full system onto three modes, which can be used to search for the best
parameter ranges. In Section \ref{sec:EP2s} we show the appearance of
continuously distributed exceptional points in the delta-functions model. The
central results are summarised in Section \ref{sec:conclusion}.

\section{Three optical wave guides with a complex
  $\mathcal{PT}$-symmetric refractive index profile}
\label{sec:system}

\begin{figure}
  \centering
  \includegraphics[width=\linewidth]{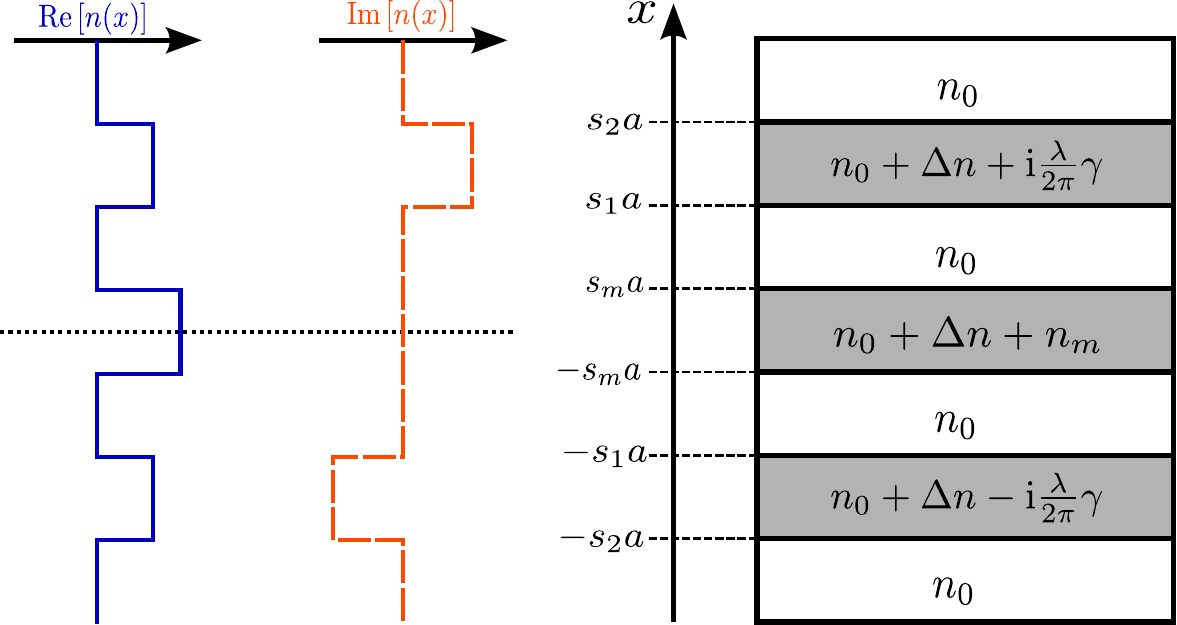}
  \caption{$\mathcal{PT}$-symmetric wave guide setup allowing for the
    occurrence of an EP3. Three coupled slab wave guides are formed on a
    background material with refractive index $n_0=3.3$ of GaAs via an index
    variation of $\Delta n = 1.3\times 10^{-3}$. The middle wave guide can
    exhibit an additional index shift given by the value of $n_m$. A gain-loss
    profile is introduced via the imaginary refractive index part  $\gamma$.
    All distances are measured in terms of a constant length scale
    $a=\SI{2.5}{\micro\m}$ via the dimensionless parameters $s_m$ and $s_{1,2}$. 
    The variation of the refractive index is only in $x$ direction and obeys
    $\mathcal{PT}$ symmetry.}
  \label{fig:system}
\end{figure}

The starting point of our investigation is the $\mathcal{PT}$-symmetric
optical wave guide system introduced in \cite{Schnabel2017a}. It consists of
three coupled planar wave guides on a background material of GaAs, which has
a refractive index of $n_0 = 3.3$ at the vacuum wavelength used in that study.
The refractive index profile is supposed to be $\mathcal{PT}$-symmetric, i.e.\
it possesses a symmetric real part representing the index guiding profile and
an antisymmetric imaginary part describing the gain-loss structure, i.e.\
$n(x)=n^*(-x)$. It extends the idea Klaiman et al.\ pursued for two wave
guides. The physical parameters consist of dimensionless scaling factors
$s_m$ and $s_{1,2}$ used to define distances in units of a constant length
$a=\SI{2.5}{\micro\m}$ (cf.\ Fig.\ \ref{fig:system})
and variations of the refractive index. The background index is shifted by a
constant value of $\Delta n = 1.3\times 10^{-3}$, and an additional shift
$n_m$ can be applied, to the middle wave guide. The gain-loss parameter is
labelled $\gamma$. The vacuum wavelength is assumed to be $\lambda=\SI{1.55}
{\micro\m}$.
                                                            
For a wave propagation of transverse electric modes along the $z$ axis the
ansatz
\begin{equation}
  \label{eq:electric-field}
  E_y(x,z,t) = \mathcal{E}_y(x)\mathrm{e}^{\mathrm{i} (\omega t - \beta z)}
\end{equation}
with $k=2\pi/\lambda$ and the propagation constant $\beta$ can be applied, and
leads to the wave equation
\begin{equation}
  \label{eq:wave-equation}
  \left(\frac{\partial^2}{\partial x^2} +
    k^2n(x)^2\right)\mathcal{E}_y(x) = \beta^2\mathcal{E}_y(x) \; ,
\end{equation}
which is formally equivalent to the one-dimensional Schr\"odinger equation
\begin{equation}
  \label{eq:schroedinger-equation}
  \underbrace{\left(-\frac{1}{2}\frac{\partial^2}{\partial x^2} 
      + V(x)\right)}_{= H_{\text{sys}}}\psi(x) = E\psi(x)
\end{equation}
with the relations
\begin{subequations}
  \label{eq:QM-analogies}
  \begin{align}
    V(x) &= -\frac{1}{2}k^2n(x)^2\;,\\
    E &= -\frac{1}{2}\beta^2\;.
  \end{align}
\end{subequations}
Thus, the formalism of $\mathcal{PT}$-symmetric quantum mechanics can be used
for the setup.

\section{Mapping onto a matrix model}
\label{sec:matrixmodel}

In the first approximation we map the full Hamiltonian of the setup shown in
Figure \ref{fig:system} to a three-mode matrix model. By this approach we
check whether the system can give rise to a third-order
branch point in a simple manner. This is clearly the case if the resulting
Hamiltonian is of the form proposed in~\cite{Graefe12}, viz.\
\begin{eqnarray}
  \label{eq:hamiltonian-graefe}
  \hat{H}_{\text{math}} = 
  \left(\begin{array}{ccc}
      a-2\mathrm{i}\gamma & \sqrt{2}v & 0 \\
      \sqrt{2}v & 0 & \sqrt{2}v \\
      0 & \sqrt{2}v & b + 2\mathrm{i}\gamma
    \end{array}\right)
\end{eqnarray}
with $\gamma,v\in\mathbb{R}$ and $a,b\in\mathbb{C}$. The real and
imaginary parts of the diagonal elements simulate the refractive index as well
as the gain (loss) behaviour in the wave guides. The parameter $v$ represents
a coupling between neighbouring wave guides via evanescent fields, and is
thus related to the distance between them. All in all $\hat{H}_{\text{math}}$
reflects a situation in which each wave guide supports a single mode.

However, this matrix is merely an abstract mathematical model without direct
connection to an experimental realisation. To establish such connection we
use the formal analogy between the wave equation \eqref{eq:wave-equation} and
the one-dimensional Schr\"odinger equation \eqref{eq:schroedinger-equation} and
calculate a matrix representation of our system in terms of
\begin{equation}
  \label{eq:matrix-mapping}
  \hat{H}' = \langle \psi_i | H_{\text{sys}} | \psi_j \rangle \;.
\end{equation}
For this purpose we assume the same real index difference $\Delta n =
1.3\times 10^{-3}$ (and $n_m = 0$) as well as the same width
$w=2a=\SI{5.0}{\micro\m}$ for all three wave guides and use the ground state
modes of each single potential well with corresponding basis functions
$\psi_{i}, \psi_{j}$. This results in a matrix of the form 
\begin{eqnarray}
  \label{eq:first-matrix}
  \hat{H}' =
  \left(\begin{array}{ccc}
      \alpha + \mathrm{i}\eta & \sigma' & \xi \\
      \sigma' & \alpha & \sigma' \\
      \xi & \sigma' & \alpha -\mathrm{i}\eta
    \end{array}\right)
\end{eqnarray}
with $\alpha,\eta,\xi\in\mathbb{R}$ and $\sigma'\in\mathbb{C}$. The use
of a common width is compatible only with $s_m=1.0$ and $s_2-s_1=2.0$, which
implies that the matrix elements still depend on the doublet $(\gamma,s_1)$,
and thus on the distance $s_1-s_m$ between the wave guides.

\begin{figure}
  \centering
  \includegraphics[width=\linewidth]{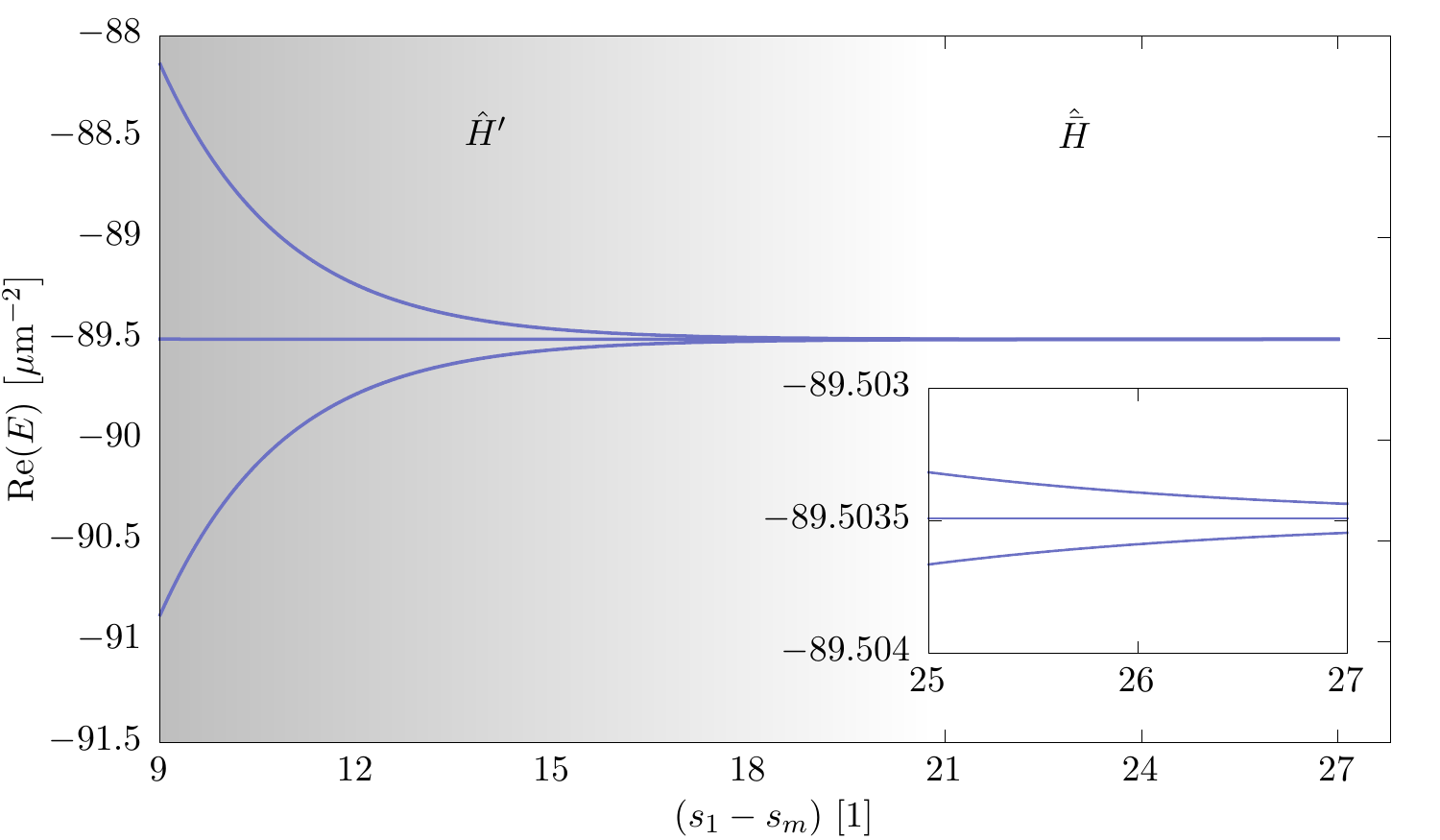}
  \caption{Real eigenenergies $E=-\beta^2/2$ of the Hamiltonian $\hat{H}'$
    as a function of the scaled distance $s_1-s_m$ between the middle wave
    guide and the outer ones. As there is a small separation between the wave
    guides there is comparatively strong coupling among the modes (grey area).
    With increasing distance the coupling becomes negligible and the system
    is well described by the new Hamiltonian $\hat{\bar{H}}$ from
    Eq.~(\ref{eq:second-matrix}).}
  \label{fig:matrix-model-distance}
\end{figure}

Eq.~(\ref{eq:first-matrix}) does not show the form predicted for the appearance
of an EP3 from Eq.~(\ref{eq:hamiltonian-graefe}) as $\xi\neq 0$ and
$\sigma'\in\mathbb{C}$. This corresponds to a situation in which also the two
outer wave guides are connected by a coupling of their waves. This can happen
due to a too small separation between the wave guides. However, for a
sufficiently large separation $s_1-s_m$ the Hamiltonian $\hat{H}'$ reduces to
the form
\begin{eqnarray}
  \label{eq:second-matrix}
  \hat{\bar{H}} = 
  \left(\begin{array}{ccc}
      \alpha + \mathrm{i}\eta & \sigma & 0 \\
      \sigma & \alpha & \sigma \\
      0 & \sigma & \alpha - \mathrm{i}\eta
    \end{array}\right)
\end{eqnarray}
with $\alpha,\eta,\sigma\in\mathbb{R}$, which resembles the desired form of
Eq.~(\ref{eq:hamiltonian-graefe}) for $a=b=0$ up to a constant shift $\alpha$.
The transition $\hat{H}'\to\hat{\bar{H}}$ can be observed in
Figure \ref{fig:matrix-model-distance}, where the real eigenenergies of
Eq.~(\ref{eq:first-matrix}) are shown as a function of the wave guides'
distances.
For comparatively small distances the energies are far apart from each
other due to a stronger coupling between the modes. In this range only
$\hat{H}'$ describes the full system correctly. For larger distances the
energies approach each other, which means that we likewise obtain an accurate
description of the system in terms of the Hamiltonian $\hat{\bar{H}}$. This is
the regime, in which a search for an EP3 is most promising.

\begin{figure}[t]
  \centering
  \includegraphics[width=\linewidth]{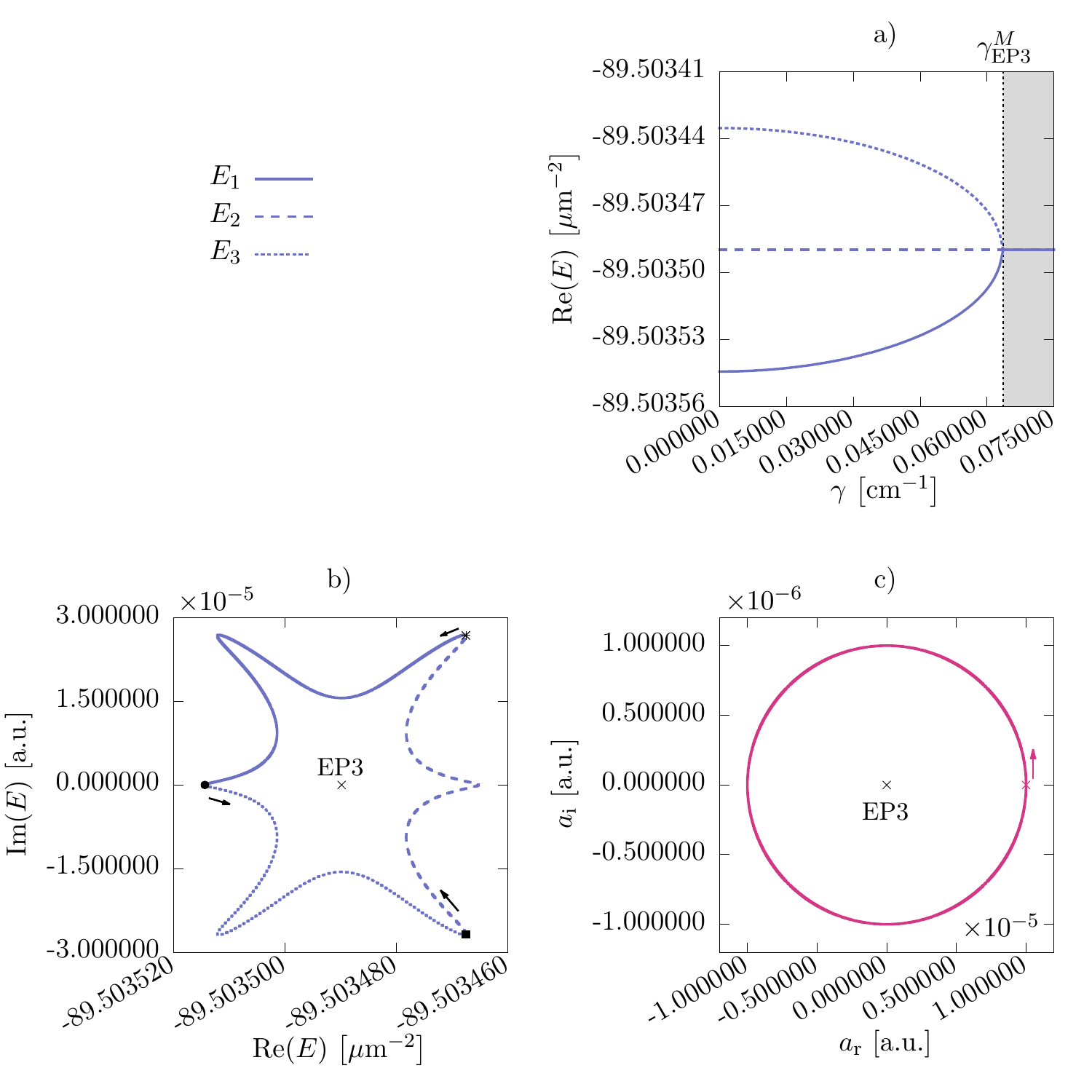}
  \caption{Evidence of a third-order exceptional point in the matrix
    model for a large separation between the wave guides ($s_m = 1.0$,
    $s_1 = 28.0$, $s_2 = 30.0$). The real index difference to the
    background material ($\Delta n = 1.3\times 10^{-3}$) was set to be
    equal for all three wave guides, i.e.\ $n_m = 0$. The coalescence of
    the three eigenenergies as a function of the non-Hermiticity
    parameter $\gamma$ is depicted in~a) and appears at
    $\gamma_{\mathrm{EP3}} = 0.0636\,\mathrm{cm}^{-1}$. The lower
    panel shows the characteristic state permutation of an
    EP3~b) as one performs a closed loop around it in a
    specific parameter space~c). Here we circle the EP3
    counter clockwise in the complex plane of the asymmetry parameter
    $a = a_{\mathrm{r}} +\mathrm{i} a_\mathrm{i}$, where the specific symbols mark
    the starting points of the permutation and the arrows the
    corresponding direction.}
  \label{fig:matrix-model-E-gamma-cycle}
\end{figure}

Knowing the shape of an appropriate wave guide we now focus on a large
separation between the wave guides with $s_m=1.0$, $s_1=28.0$ and $s_2=30.0$
and verify the existence of the EP3. We find it in the spectra by varying the
gain-loss coefficient $\gamma$. The result is depicted in
Figure \ref{fig:matrix-model-E-gamma-cycle}~a).
Obviously we obtain the coalescence of all three real eigenenergies for
$\gamma_{\mathrm{EP3}}^M=0.0636\,\mathrm{cm}^{-1}$ in a cubic root branch point
singularity. Beyond this point the spectrum becomes complex with two complex
conjugate energies and one with vanishing imaginary part. Note that the middle
state stays widely unaffected by an increase of the non-Hermiticity parameter,
wich is also found in the more realistic descriptions \cite{HeissWunner16,%
  Schnabel2017a} as well as flat band systems \cite{Ge2015a,Ge2017a}.

To ascertain that this is indeed an EP3 we follow a standard procedure and
perform a closed loop in a suitable parameter space around the supposed branch
point singularity. To do so we have to introduce the complex parameters $a,b$
from Eq.~(\ref{eq:hamiltonian-graefe}) breaking the underlying $\mathcal{PT}$
symmetry. We restrict ourselves to the specific choice $a=b=a_{\mathrm{r}}
+\mathrm{i} a_\mathrm{i}$ and add this to $\hat{\bar{H}}$, ending up with
\begin{eqnarray}
  \label{eq:matrix-asymmetry}
  \hat{H}_A = \hat{\bar{H}} + a
  \left(\begin{array}{ccc}
      1 & 0 & 0 \\
      0 & 0 & 0 \\
      0 & 0 & 1
    \end{array}\right)
\;.
\end{eqnarray}
Using this form the circle is performed in the complex plane of $a$ along
an ellipse with the parametrization
\begin{eqnarray}
  \label{eq:parametrization-ellipse-matrix}
  \left[0,2\pi\right]\to\mathbb{R}^2\,,\,\varphi\mapsto
  \left(\begin{array}{c}
      a_{\mathrm{r}} \\
      a_\mathrm{i}
    \end{array}\right)
    =
    \left(\begin{array}{c}
        10^{-5}\cos\varphi\\
        10^{-6}\sin\varphi
      \end{array}\right)
\end{eqnarray}
as illustrated in Figure \ref{fig:matrix-model-E-gamma-cycle}~c). The
corresponding state permutation is depicted in
Figure \ref{fig:matrix-model-E-gamma-cycle}~b) and clearly exhibits
the threefold exchange behaviour of an EP3.

\section{Continuously distributed exceptional points around
  the EP3}
\label{sec:EP2s}

The third-order exceptional point investigated in Section \ref{sec:matrixmodel}
was verified with a parameter space circle and its typical threefold
permutation behaviour. As was found in \cite{Schnabel2017a} this can become
a difficult task in a realistic setup. On the one hand, not every parameter
space circle leads to the threefold permutation. A twofold state exchange
misleadingly indicating an EP2 is also possible for certain parameter choices
\cite{Graefe12,Heiss2016}. On the other hand, additional exceptional points,
which are accidentally located within the area enclosed by the parameter
space loop, can distort the signature of the EP3. In the spatially extended
investigation of \cite{Schnabel2017a} it turned out that the EP3 is
accompanied by continuous distributions of exceptional points in such a way
that it is very hard to find a parameter plane, in which a circle reveals
the pure cubic-root branch point signature of the EP3. Here we show that this
is not only a property of the special shape of the three wave guides used in
\cite{Schnabel2017a} but a generic feature of three coupled guiding profiles.
To do so, we return to the delta-functions model from \cite{HeissWunner16}.

The model is given by an effective
Schr\"odinger equation of the form
\begin{align}
  \label{eq:SG-deltasystem}
  -\psi''(x)&-\bigl[(1+\mathrm{i}\gamma)\delta (x+b) + \Gamma\delta (x) 
  \notag\\&\qquad + (1-\mathrm{i}\gamma)\delta (x-b)\bigr]\psi(x)
  = -k^2\psi(x) \; ,
\end{align}
where three delta-function potential wells are located at $x=\pm b$ and
$x=0$. Loss is added to the left well and the same amount of gain is added
to the right one via the parameter $\gamma$. The units are chosen in such a way
that the strength of the real and imaginary parts of the two outer wells is
normalised to unity, while in the middle well we allow for a different depth
given by the real parameter $\Gamma > 0$, similarly to its spatially extended
counterpart from Section \ref{sec:system}. As the system is non-Hermitian 
the eigenvalues $k$ are complex in general with $\mathrm{Re}(k) >0$. We are
interested in bound state solutions with real eigenvalues, and the bound state
wave functions have the form
\begin{align}
  \label{eq:wavefunc-deltasystem}
  \psi(x) &=
  \begin{cases}
      A\mathrm{e}^{kx} &:\; x < -b \\
      2(r\cosh(kx) + \varrho_1\sinh(kx)) &:\; -b<x<0 \\
      2(r\cosh(kx) + \varrho_2\sinh(kx)) &:\;0<x<b \\
      B\mathrm{e}^{-kx} &:\; x > b
    \end{cases}
    \;.
\end{align}
As the continuity conditions and the discontinuity conditions for the wave
functions and their first derivatives have to be fulfilled at the delta
functions we obtain a system of linear equations in the form
\begin{align}
  \label{eq:linear-eqs-deltasystem}
  \mathcal{M}
  \begin{pmatrix}
      r \\
      \varrho_1 \\
      \varrho_2
    \end{pmatrix}
  &= \mathbf{0}\;,
\end{align}
for which nontrivial solutions exist if the corresponding secular equation
\begin{align}
  \label{eq:secular-eq}
  \mathrm{det}(\mathcal{M}) &= 
  \Gamma\bigl(\mathrm{e}^{-4kb}(1+\gamma^2) - 2\mathrm{e}^{-2kb}(\gamma^2-2k+1) \notag\\ &\qquad + \gamma^2 + (2k-1)^2\bigr) +2k\bigl(\mathrm{e}^{-4kb}(1+\gamma^2) \notag\\ &\qquad- \gamma^2 - (2k-1)^2\bigr) = 0
\end{align}
vanishes. Hence we obtain the eigenvalues $k$ as roots of the
determinant $\mathrm{det}[\mathcal{M}](k)\equiv f(k)$ depending on the
distance $b$, the non-Hermiticity parameter $\gamma$, and the
parameter $\Gamma$ of the middle well. Assuming $k$ to be purely real,
the position of the third-order branch point singularity is fixed by
\begin{equation}
  \label{eq:cond-EP-deltasystem}
  f(k) = \frac{\partial f}{\partial k} = \frac{\partial^2 f}{\partial k^2}
  = 0\;.
\end{equation}
With $\Gamma_{\mathrm{EP3}} = 1.002$ the EP3 appears at
\begin{subequations}
  \label{eq:EP-params-deltasystem}
  \begin{align}
    \gamma_{\mathrm{EP3}}^\delta &= 0.06527796794065678\;,\\
    b_{\mathrm{EP3}} &= 6.2012417361076206 \;,\\
    k_{\mathrm{EP3}} &= 0.49584858490334327\;.
  \end{align}
\end{subequations}
This can be verified by circling this point in the complex plane of
the distance $b$ (as it was done in \cite{HeissWunner16}) or by
introducing asymmetry parameters breaking the underlying $\mathcal{PT}$ 
symmetry.

\begin{figure}
  \centering
  \includegraphics[width=\linewidth]{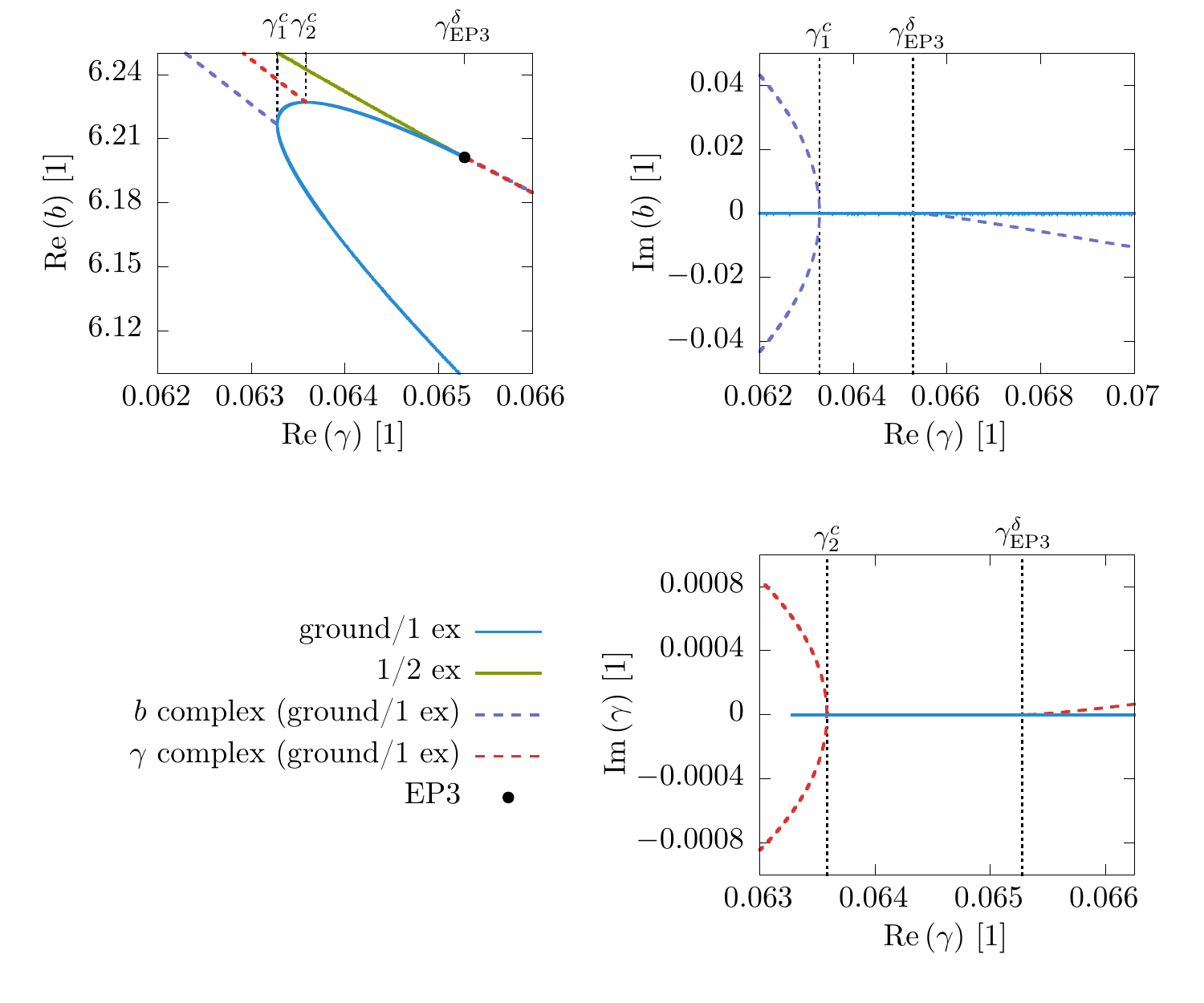}
  \caption{Continuously distributed second-order exceptional points in
    the $b$-$\gamma$ space of the idealised system made up of three
    delta-functions potentials. The system's EP3 appears at
    $\gamma_{\mathrm{EP3}}^\delta$, from which several EP2s arise (top
    left, solid lines) either between the ground and first excited
    (1.ex) or between the first and second excited (2.ex) state. They
    reveal an even more profound branch structure at the points
    $\gamma_1^c$ and $\gamma_2^c$, which can be explained in terms of
    an analytical continuation of $b$ or $\gamma$ (top right and
    bottom right).}
  \label{fig:EP2-deltasystem}
\end{figure}

A verification without $\mathcal{PT}$ symmetry breaking, i.e.\ a circle around
the EP3 in the $b$-$\gamma$ space turns out to be impossible in this
simplified model as it is always dominated by a signature belonging to an EP2.
This suggests the conclusion that in analogy with the spatially extended model
from \cite{Schnabel2017a} the EP3 may be accompanied by EP2s, which
disturb the exchange behaviour. To expose this behaviour we attenuate the
condition of Eq.~(\ref{eq:cond-EP-deltasystem}) to a twofold zero, from which
we get the pair of variables $(k,\gamma)$ or $(k,b)$ and therefore the
positions of the EP2s via a two-dimensional root search. The results are shown
in Figure \ref{fig:EP2-deltasystem} (top left). It can be seen that the
EP2s are distributed continuously around the EP3 in the $b$-$\gamma$ space.
The lines represent either EP2s between the ground state and the first
excited mode or between both excited modes. They coalesce at the
position of the EP3 at $\gamma_{\mathrm{EP3}}^\delta$ leaving a knee
in the parameter space. Moreover there appear more branches at
$\gamma_{1,2}^c$ along the blue line, which cannot be explained by
purely real parameters $b$ and $\gamma$. Hence we either continue $b$
or $\gamma$ analytically into the complex plane and allow for $k\in\mathbb{C}$,
which turns the two-dimensional root search into a four-dimensional one. The
resulting effects on the $\mathrm{Re} (\gamma)$-$\mathrm{Im} (b)$ space or
$\mathrm{Re} (\gamma)$-$\mathrm{Im} (\gamma)$ space are depicted on the
right-hand side of Figure \ref{fig:EP2-deltasystem}.

\section{Conclusion}
\label{sec:conclusion}

In this paper we applied two approximations to the system of three coupled
$\mathcal{PT}$-symmetric wave guides studied in \cite{Schnabel2017a}. In a
mapping of the system to a three-mode matrix model we could show that the matrix
can serve as an intuitive guide to parameter regimes, in which the prospects
of finding a third-order exceptional point are best. This is exactly the
case when the correctly mapped matrix assumes, due to appropriately chosen
physical parameters, the shape proposed in \cite{Graefe12}.

The continuous distributions of EP2s around the EP3 of interest in the space
of the accessible physical parameters can be found in the much simpler
delta-functions model from \cite{HeissWunner16}. Thus, it is possible to
search for adequate physical parameters allowing for the identification of
the EP3 via its characteristic threefold state permutation without the need
of having to solve the full problem.

In principle the approach of extending the system with additional wave guides
to allow for higher-order exceptional points can be continued. With four wave
guides it should for example be possible to access a fourth-order exceptional
point. The observations made in this work suggest that all further extensions
should first be studied in simple approaches before a laborious modelling
of a realistic physical setup is done. An $N$-mode matrix model can tell
whether a promising search for an EP$N$ in a setup with $N$ wave guides is
possible. If this is the case, it can provide rough estimates for suitable
physical parameters.

The reduction of the full system to delta functions leads to much simpler
equations but preserves the whole richness of effects. As such it can be used
as a first access to the structure of the eigenstates. In particular, it can
be used to evaluate whether an identification of the EP$N$ via the $N$-fold
permutation of the eigenstates seems to be feasible. This can give valuable
information before costly numerical calculations of the full system are
done.

\begin{acknowledgements}
  GW and WDH gratefully acknowledge support from the National Institute for
  Theoretical Physics (NITheP), Western Cape, South Africa. GW expresses his
  gratitude to the Department of Physics of the University of Stellenbosch
  where parts of this paper were developed.
\end{acknowledgements}

\bibliographystyle{actapoly}

\end{document}